\documentclass[a4paper]{jpconf}
\usepackage{graphicx}
\usepackage{harvard}
\begin{document}
\title{The galactic population of white dwarfs}

\author{Ralf Napiwotzki}

\address{Centre for Astrophysics Research, University of Hertfordshire, College Lane, Hatfield AL10~9AB, UK}

\ead{r.napiwotzki@herts.ac.uk}

\begin{abstract}
The contribution of white dwarfs of the different Galactic
populations to the stellar content of our Galaxy is only poorly
known. Some authors claim a vast population of halo white dwarfs, which would
be in accordance with some investigations of the early phases of
Galaxy formation claiming a top-heavy
initial--mass--function.  Here, I present a model of the population of
white dwarfs in the Milky Way based on observations of the local white
dwarf sample and a standard model of Galactic structure. This model will be
used to estimate the space densities of thin disc, thick disc and halo
white dwarfs and their contribution to the baryonic mass budget of the Milky
Way. One result of this investigation is that white dwarfs of the halo
population contribute a large fraction of the Galactic white dwarf number
count, but they are not responsible for the lion's share of
stellar mass in the Milky Way. Another important result is the
substantial contribution of the -- often neglected -- population of 
thick disc white dwarfs. Misclassification of
thick disc white dwarfs is responsible for overestimates of the halo
population in previous investigations.
\end{abstract}

\section{Introduction}

The stellar populations of our Galaxy are divided into the thin disc,
thick disc, halo and Galactic bulge. Only the thin disc still produces
young stars. Star formation has ceased in all other populations at
least 10\,Gyr ago and all stars born more massive than the Sun have
already evolved away from the main sequence, producing faint remnants:
mostly white dwarfs plus some neutron stars and black holes. If the
initial--mass--function in the early phases of galaxy formation was as
top heavy as claimed by some high redshift investigations (e.g.\
\citename{B.L.F2005}\ 2005) white dwarfs could be responsible for an
important fraction of the Galactic mass budget.

Observations by the MACHO project of microlensing events in the
Magellanic Clouds attributed to white dwarfs \cite{A.A.A2000} and a
deep proper motion survey for high-velocity white dwarfs by
\citeasnoun{O.H.D2001} appeared to confirm a high density of white
dwarfs in the Galactic halo. Both results remain controversial.  A
recent analysis and discussion of the MACHO results is provided by
\citeasnoun{T.C.I2008}. The \citeasnoun{O.H.D2001} results have been
criticised by \citeasnoun{R.S.H2001} and \citeasnoun{P.N.A2003}, but
open questions remain. Space densities of thick disc and halo white
dwarfs and their contribution to the Milky Way mass budget are still
open questions.

I will present a Monte Carlo simulation of 
the white dwarf populations in our Galaxy. The simulation models the complete
Galactic content of white dwarfs and post-AGB stars. Simulated stars can be
selected according to criteria resembling the selection of observed samples. 
After a short description of the model I will describe how it is 
calibrated using work based on observations
of white dwarfs with the Supernova type Ia Progenitor surveY (SPY;
\citename{N.C.D2003}\ 2003). This population model will be used to
estimate the local space density of the three white dwarf populations
and their contribution to the Galactic mass. Sect.~4 compares results
with the \citeasnoun{O.H.D2001} observations. This contribution
finishes with a discussion
in Sect.~5.

\section{Population model and calibration}
\label{s:popmod}

The simulations are based on the model of the Galactic structure
presented by \citeasnoun{R.R.D2003}. All four stellar populations,
thin disc, thick disc, halo and bulge (the latter not of relevance
here) are included.  Extinction by the gas and dust layer in the
Galactic plane is taken into account using a simple analytical
model. The impact of extinction on the white dwarf simulations
presented here is not very large. Density laws, but not the zero
points, and velocity dispersions were taken from
\citeasnoun{R.R.D2003}. Different from most white dwarf population
synthesis models the age dependent scale height and velocity
dispersion of the thin disc is taken into account. Stars are created
at random positions according to the density laws, a random age and a
random mass drawn from a Salpeter initial--mass--function. Short lived
star bursts early in the lifetime of our Galaxy are assumed for halo
and thick disc. The thin disc is modelled with a constant star
formation rate.

The white dwarf progenitor lifetime is computed from the grids of evolutionary
models created by the Padova group \cite{G.B.B2000}. The simulations
presented here used one single metallicity for every
population. More realistic metallicity distributions are
implemented now. The standard model uses the initial--final--mass
relation (IFMR) from \citeasnoun{Wei2000} and the cooling tracks are a
blend of the \citeasnoun{Blo1995b} post-AGB tracks for hot white
dwarfs and white dwarf sequences computed by \citeasnoun{C.B.A2000}. Other
IFMRs (e.g., that proposed by \citename{D.N.B2006}\ 2006) and
\citeasnoun{F.B.B2001} cooling tracks are currently implemented and
their effect will be tested.

A direct census of white dwarfs in the Galactic halo is impossible so far.  All
substantial samples of field white dwarfs are essentially collected from the
local volume.  Even at $V=20$ (roughly corresponding to the
spectroscopic white dwarf sample extracted from the SDSS archives) more than
98\% of the observed white dwarfs are closer than 1\,kpc in a region still dominated by the thin disc. However, classification of white dwarf
membership in the Galactic populations is possible based on kinematic
criteria. \citename{P.N.A2003}\ (2003, 2006) combined accurate radial
velocity measurements of white dwarfs from the SPY sample 
(\citename{N.C.D2001}\ 2001, 2003)
with proper motion data from recent catalogues (UCAC2,
USNO-B, SuperCosmos; see \citename{P.N.H2006}\ 2006 for details)
supplemented by own measurements. Distances were estimated from the
spectroscopic analysis of the SPY spectra (\citename{K.N.C2001}\ 2001,
\citename{V.K2005}\ 2005). This work resulted in a sample of 400 white dwarfs
brighter than $V=16.5$ for which the complete 6-dimensional set of
parameters in phase space (spatial and kinematical) is available.

\citename{P.N.A2003}\ (2003, 2006) developed criteria to assign
population memberships based on kinematical properties. An important
tool for this analysis was the position in the $U-V$ plane (examples
are shown in Fig.~\ref{f:UV}). $U$, $V$ and $W$ are the velocity
components in the Galactic reference system. $U$ is the component in
direction of the Galactic centre, $V$ the tangential component in
direction of the orbit around the centre and $W$ the velocity
perpendicular to the Galactic plane. The classification was further
refined using the position in the angular momentum--eccentricity
plane and a check of the Galactic orbit as computed with the
\citeasnoun{O.B1992} code. While thin disc stars have low eccentricity
orbits confined to small distances from the Galactic plane, most thick
disc stars have orbits of somewhat larger eccentricity reaching larger
distances. Halo stars do not take part in the rotation of the Galactic
disc around the centre and can have orbits which reach very large
distances from the plane, sometimes with extreme eccentricities.
\citename{P.N.A2003}\ (2003, 2006) calibrated the classification
criteria with main sequence stars, for which abundance information was
available to assess population memberships. The resulting limits in
the $U-V$ plane are shown in Fig.~\ref{f:UV}.

\section{Galactic white dwarf populations}

\citeasnoun{P.N.H2006}\ classified a total of 398 white dwarfs using
the criteria outlined above. The results are given in
Table~\ref{t:pauli}. We have to keep in mind that the numbers apply to
a {\it brightness limited} sample ($V\le 16.5$) of local white
dwarfs. Three more steps are necessary before the white dwarf
contributions to the Galactic mass budget can be estimated.

\begin{table}[!ht]
\caption{Population classification of 398 white dwarfs from the SPY sample by
\citeasnoun{P.N.H2006}.}
\label{t:pauli}
\begin{center}
\begin{tabular}{lrr}\br
population   &$N_{\mathrm{WD}}$ &\\ \mr
thin disc    &368   &92\%\\
thick disc   &23    &6\%\\
halo         &7     &2\%\\\br
\end{tabular}
\end{center}
\end{table}

\subsection{Correction for contamination} 
\begin{figure}[!ht]
\includegraphics[,angle=-90,width=8cm]{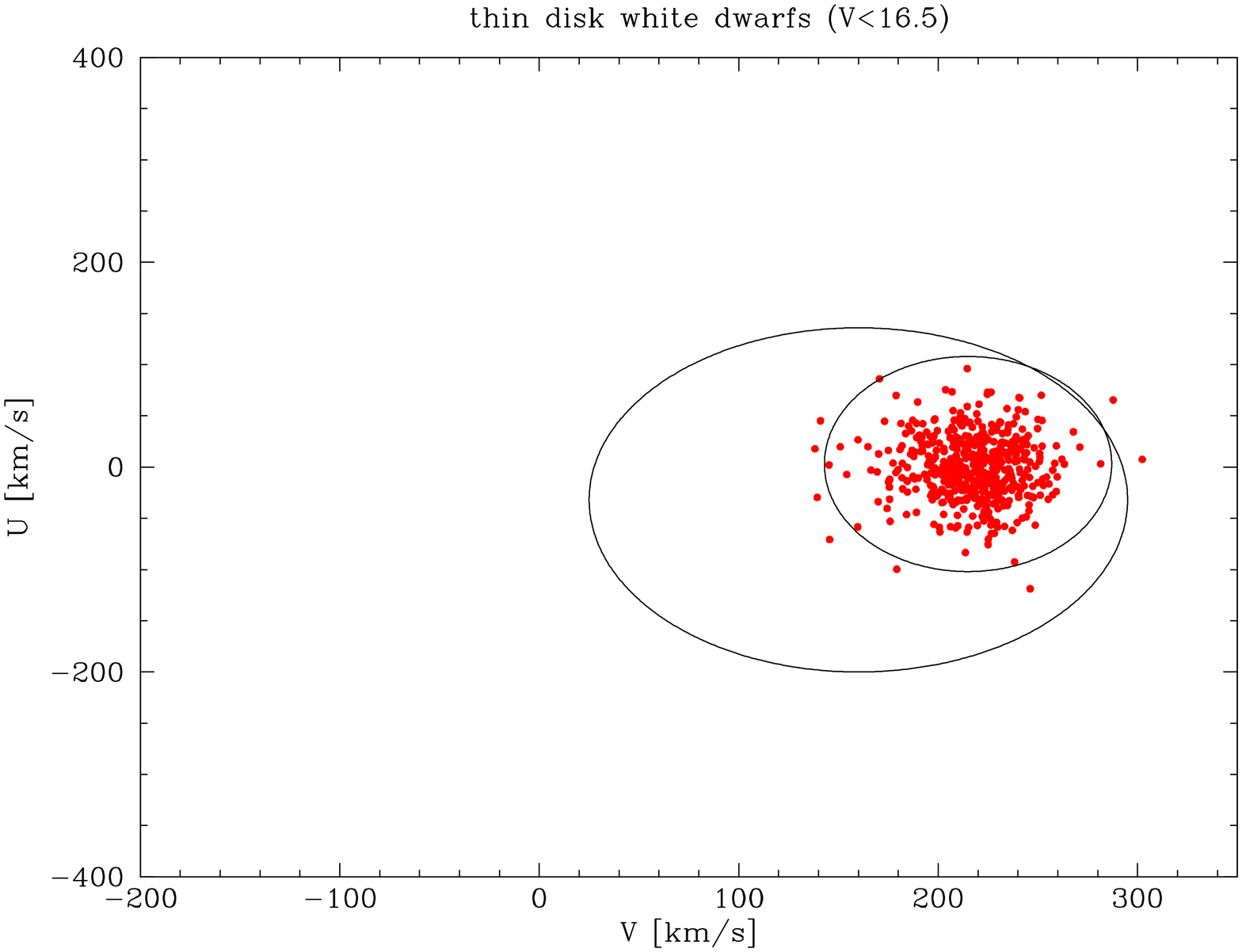}\hspace{3mm}
\includegraphics[,angle=-90,width=8cm]{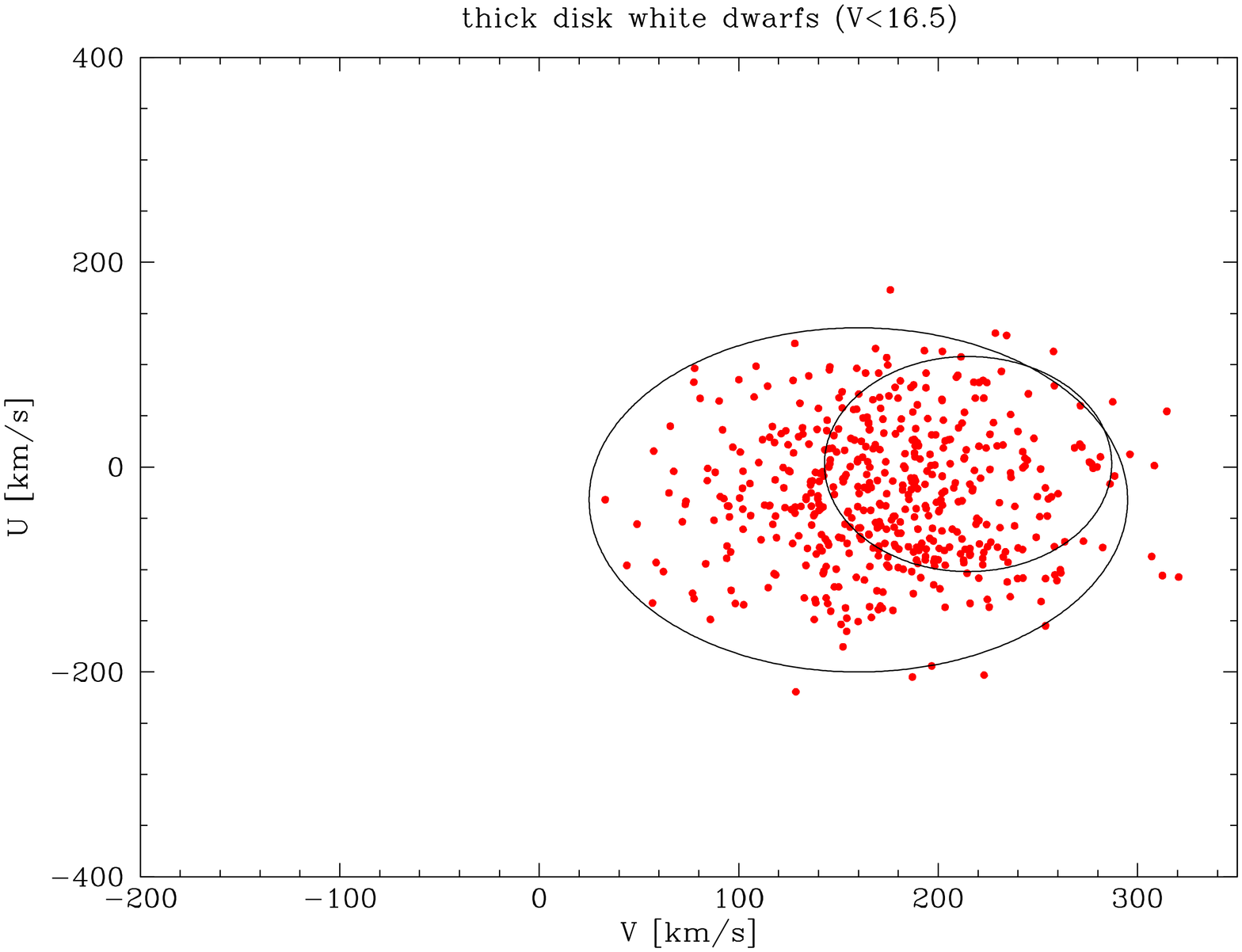}\\[2mm]
\includegraphics[,angle=-90,width=8cm]{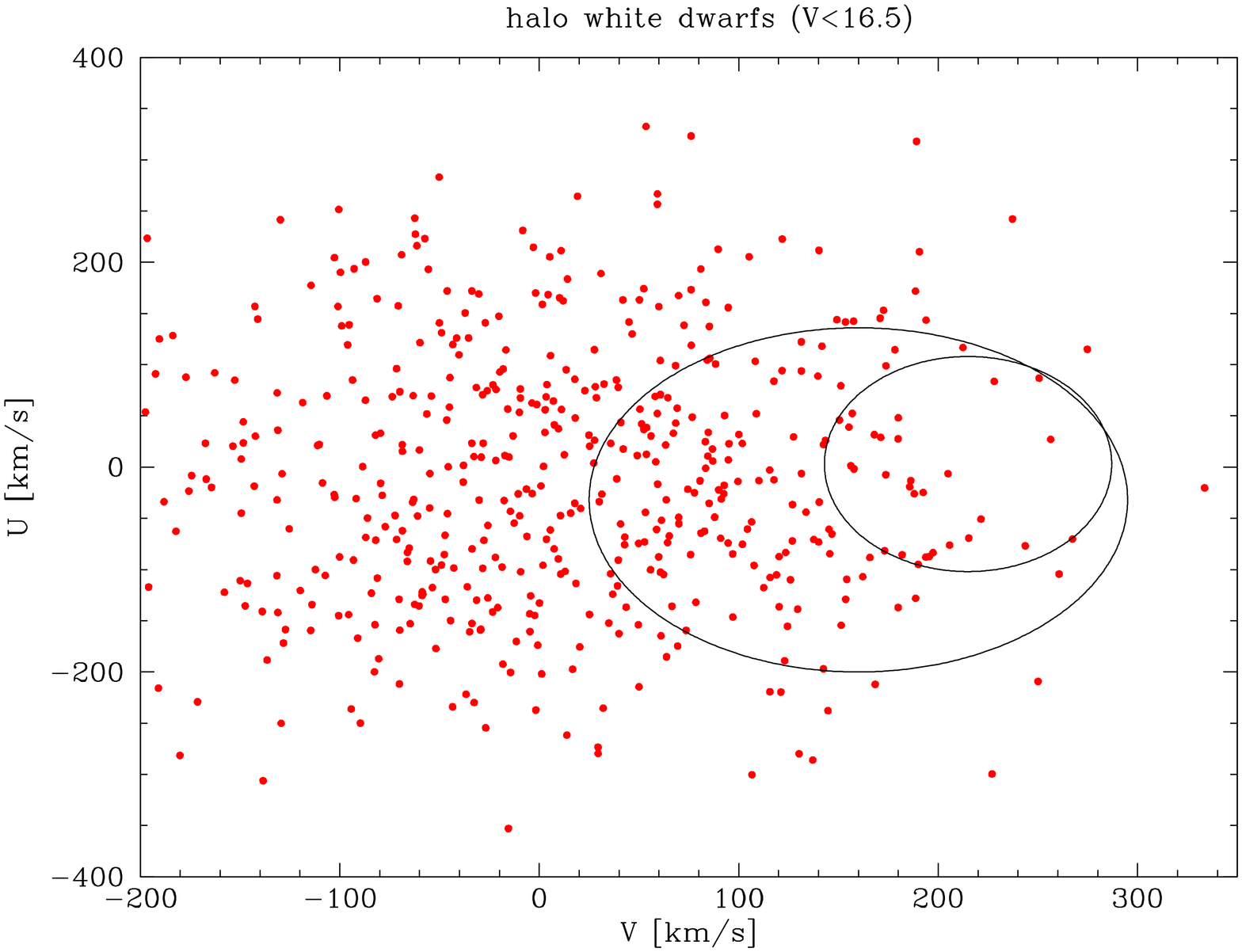}\hspace{3mm}
\begin{minipage}[t]{7cm}
\hspace{1cm}\\[2cm]
\caption{Simulated white dwarf samples in the $U-V$ plane. The smaller
ellipse indicates the $3\sigma$ limit for the thin disc and the larger
one the $3\sigma$ limit for the thick disc.}
\label{f:UV}
\end{minipage}
\end{figure}

It is inevitable that, say, some
thick disc stars have kinematical parameters which are typical for
thin disc white dwarfs. This is not due to errors in the classification
criteria, but an intrinsic property of the populations. This is
demonstrated with simulated white dwarf samples in the $U-V$ plane in
Fig.~\ref{f:UV}. 

The first plot shows a simulated thin disc sample. The vast majority
of stars lies within the thin disc boundaries defined by
\citeasnoun{P.N.H2006}. Only a few white dwarfs are scattered into the
thick disc or halo regions.  Note that the kinematic properties of the
white dwarfs in the simulation are independent of the
\citename{P.N.H2006}\ calibration, providing an independent
verification of the chosen boundaries. The second plot shows the
simulated thick disc sample. Again the outer boundaries are well
selected, but it is obvious that quite a few thick disc white dwarfs
will be misclassified as members of the thin disc. The third plot
shows the result for the halo simulation. As expected the halo white
dwarfs are distributed all over the $U-V$ plane, including the regions
defined for the thin and thick disc.

The Monte
Carlo model of the white dwarf populations can be used to correct for
contamination or deficit. Since the corrections depend on the adopted
space densities and the SPY results listed in
Table~\ref{t:pauli} are used to calibrate the space densities in the
population model, a few iterations are necessary until self-consistency
of the corrected numbers is achieved. The final results are shown in
Table~\ref{t:correct}. The main change to the uncorrected numbers is 
a substantial increase of the thick disc contribution. Corrections for 
the halo population cancel out.

\begin{table}[!ht]
\caption{Corrected results for white dwarf population membership. The
column labelled SPY repeats the uncorrected fractions in the SPY
sample given in Table~\ref{t:pauli}.  The next one (corr.) shows the
fractions after correction for misclassifications. Relative fractions
of the volume limited sample and local space densities $\rho$ are
given in the following columns. The last two columns give the relative
fraction and total number extrapolated for the complete Galaxy.}
\label{t:correct}
\begin{center}
\begin{tabular}{lrrrrrc}\br
population   &SPY &corr.  &\multicolumn{2}{c}{vol.\ limited} 
&\multicolumn{2}{c}{Galaxy}\\
             &       &      &     &\multicolumn{1}{c}{$\rho$ [pc$^{-3}$]} 
&  &$N/10^9$\\\mr
thin disc    &92\%   &86\%  &59\% &$2.9\times 10^{-3}$ &17\%  &2.0\\
thick disc   &6\%    &12\%  &35\% &$1.7\times 10^{-3}$ &34\%  &3.9\\
halo         &2\%    &2\%   &6\%  &$2.7\times 10^{-4}$ &49\%  &5.6\\\br
\end{tabular}
\end{center}
\end{table}

\subsection{Contributions in a volume limited sample}

\begin{figure}[!ht]
\centering
\includegraphics[width=12cm,bb=20 155 565 690]{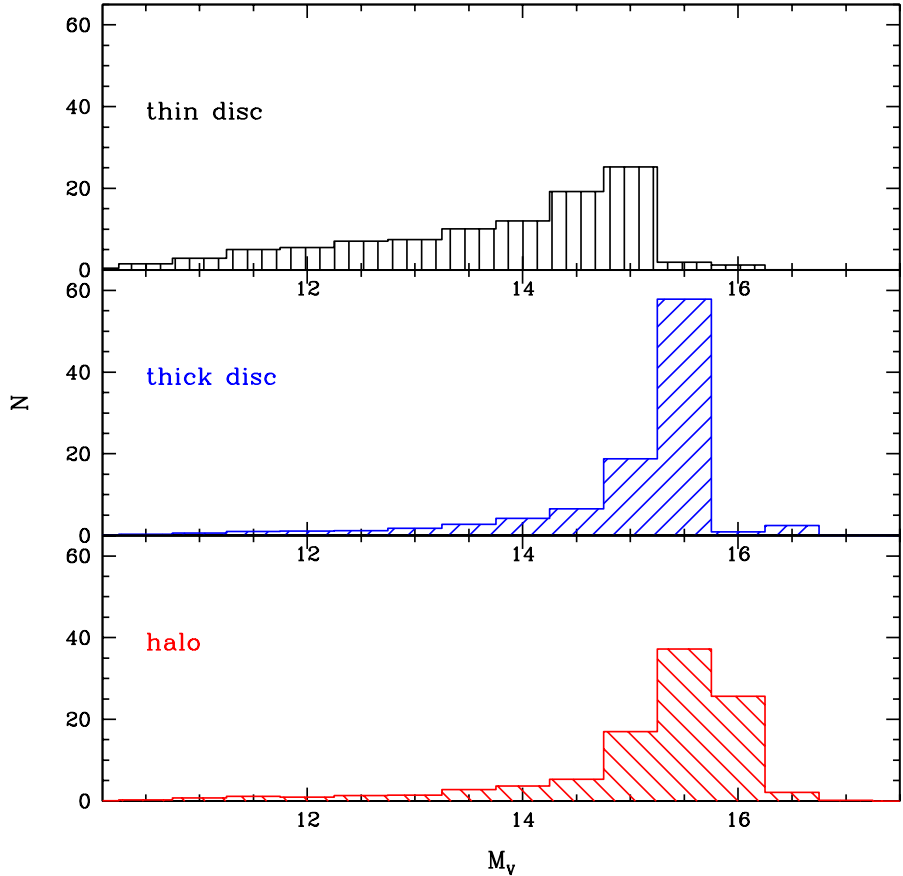}
\caption{Simulated luminosity functions of thin disc, thick disc and
halo white dwarfs. Each luminosity function is normalised to a total of 100 
white dwarfs, although the real numbers of simulated white dwarfs are much higher.}
\label{f:LF}
\end{figure}

The corrected sample is still brightness limited. Many older and
cooler white dwarfs are too faint to be observed. Since the thick disc and the
halo populations are very old, they contain many old white dwarfs of very low
luminosity. The luminosity functions in Fig.~\ref{f:LF} show the much
larger fraction of intrinsically faint white dwarfs compared to the thin
disc. This results in an overrepresentation of the thin disc in any
brightness selected sample of white dwarfs. The simulated white dwarf sample allows a
straightforward correction of this bias (Table~\ref{t:correct}). We
can use these results to compute the local space densities of the
three populations.  The local space density of all white dwarfs was estimated
by \citeasnoun{H.S.O2008} to be $\rho = 4.8\times
10^{-3}\,\mathrm{pc}^{-3}$ based on a local sample of white dwarfs ($d\le
13$\,pc). Some objects were added or removed compared to an earlier
version of the 13\,pc sample \cite{H.O.S2002} and some low proper
motion cool white dwarfs are probably still missing \cite{S.P.N2004}, but
overall this estimate of the space density appears fairly robust and
dramatic revisions are not expected for the future. The local white dwarf space
densities of the three populations in Table~\ref{t:correct} are
computed combining the simulation results with the
\citeasnoun{H.S.O2008} white dwarf space density.

\subsection{Galactic white dwarf populations}
\label{s:galpol}

The final step is the extrapolation of results to the whole
Galaxy. This can be easily done from the Monte Carlo simulations which
produce a complete census of all white dwarfs in the Galaxy. 
Most of them are discarded in subsequent
steps, because they are too distant or too faint, but the numbers
are available after a run. The estimates of total white dwarf numbers and relative
fraction of the Galactic population are given in Table~\ref{t:correct}. 
The results show that the Galactic population of white dwarfs is
dominated by thick disc and halo. However, the white dwarf
contribution to the total stellar mass of the Galaxy amounts not not
more than $\approx$10\%, far less than expected from the results of
\citeasnoun{A.A.A2000} and \citeasnoun{O.H.D2001}. 

A possible explanation for this apparent discrepancy will be discussed
in the next section. However, let me first point out the limitations
of these estimates. Obviously, the extrapolation relies on the validity
of the model of the Galactic structure by \citeasnoun{R.R.D2003}
implemented in the population model. Possible future modifications of
state--of--the--art modelling of our Galaxy could result in significant
changes of the white dwarf numbers. It can be expected that the relative
numbers of white dwarfs and the relative contribution to the Galactic stellar
mass are more robust than the total numbers. Another limitation comes
from the number statistics. While the thin disc and thick disc white dwarf
samples have a reasonable size, the size of the observed halo white dwarf
sample is only seven. This causes a considerably statistical
uncertainty. Improvement can only be expected from larger observed
samples.

\section{The Oppenheimer sample of high proper motion white dwarfs}

Using the Monte Carlo model of the Galactic white dwarf population I have
shown that the local, brightness limited sample observed by SPY
implies a modest contribution of white dwarfs to the Galactic mass budget.
This is in contrast to the claims by \citeasnoun{A.A.A2000} and
\citeasnoun{O.H.D2001} of large contributions of white dwarfs in the Galactic
halo to the total mass of our Galaxy. Can the different results be
reconciled?  The interpretation of the MACHO results reported by
\citeasnoun{A.A.A2000} experienced some revision over the years. The reader is
referred to \citeasnoun{T.C.I2008} and references therein for a discussion.
Here I will have a look at the \citeasnoun{O.H.D2001} survey. 

\citename{O.H.D2001}\ performed a survey of 4900 square degrees near
the South Galactic Cap using digitised photographic plates. They
selected high proper motion objects with
$\mu>0.33''/\mathrm{yr}$. These were plotted in a reduced proper
motion diagram and a list of 126 candidates with a position consistent
with halo membership was produced. Follow-up spec\-tro\-scopy confirmed
that most of them were white dwarfs. \citename{O.H.D2001}\ identified
38 can\-di\-dates, which they claimed were cool white dwarfs in the
Galactic halo.

I simulated a sample of white dwarfs using the local densities given
in Table~\ref{t:correct} and selection criteria similar to
\possessivecite{O.H.D2001} survey. The major difference being that the
simulations constructed the reduced proper motion diagram using
Johnson $B$ and $V$ filters instead of the photographic $B_J$ and $R$
measurements used in their investigation. I tried to define cuts
similar to those of \citename{O.H.D2001}\ in the reduced proper motion
diagram, but the match will not be perfect. Another effect not
modelled in the simulation is caused by the large scatter associated
with photographic brightnesses and colours. This can result in objects
scattered into and out of the selection regions. However, the expected
impact of both effects are relatively minor and will not affect the
qualitative interpretation of the results.

\begin{figure}[!ht]
\centering
\includegraphics[width=12cm,bb=20 155 565 690]{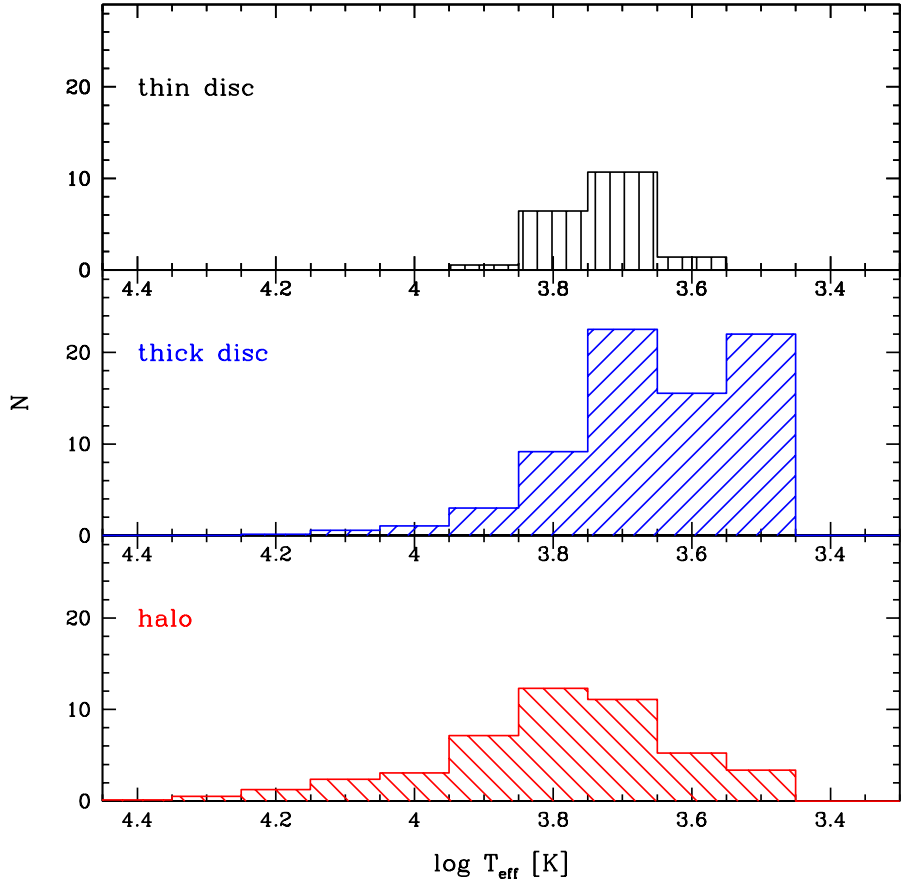}
\caption{Simulated sample of high proper motion white dwarfs analogous to the
selection criteria used by \citeasnoun{O.H.D2001}.}
\label{f:Oppen}
\end{figure}

The simulated sample consists of 140 candidates, which is consistent
with the \citeasnoun{O.H.D2001} numbers. On the observational side it
has to be expected that some candidates are missed, e.g.\ because of
blending problems. Moreover, a certain error range has to be assumed
for the SPY based calibration of white dwarf densities, as discussed
in Sect.~\ref{s:galpol}. Fig.~\ref{f:Oppen} displays the temperature
distributions of the simulated high proper motion sample. This plot
verifies \possessivecite{O.H.D2001} assumption that the thin disc
contribution is small. However, the vast majority of cool-ish high
white dwarfs in this sample belongs to the thick disc, not the halo
population.  This was already suspected by \citeasnoun{R.S.H2001}.
The temperature distributions in Fig.~\ref{f:Oppen} are somewhat
counter intuitive. As Fig.~\ref{f:LF} shows, the (simulated) halo
luminosity function contains even more faint and cool white dwarfs
than the thick disc sample and typical space velocities are
higher. However, space densities of the halo population are lower and
typical distances of white dwarfs larger. Thus, while cool halo white
dwarfs would easily make it through the proper motion criteria, many
of them are not included in the survey because they are fainter than
the brightness limit.

The misclassification of thick disc white dwarfs as halo objects is the crucial
factor causing the very high estimate for the mass of the population
of halo white dwarfs by \citeasnoun{O.H.D2001}. They overestimated the local
density of halo white dwarfs. Since the volume filled by the Galactic halo is
much larger than the volume of the thick disc, the extrapolation to
the whole Galaxy led to an dramatic overestimate of the mass. With
correct classification of the white dwarfs the discrepancy disappears and SPY
and \citename{O.H.D2001}\ sample are in good agreement.

\section{Discussion and future work}

A Monte Carlo simulation of the white dwarf population in our Galaxy
is constructed.  It was calibrated using population classifications
derived by \citeasnoun{P.N.H2006} from an analysis of the brightness
limited SPY sample. It is straightforward to model selection criteria
for observational samples in the Monte Carlo simulation. This was used
to apply a statistical correction to the SPY results and compute the
contribution of the Galactic populations to the local white dwarf
space density and the Galactic number count. The estimated fraction of
white dwarfs in the halo is $\approx$50\% of the Galactic white
dwarfs, but the overall contribution of white dwarfs to the Galactic
mass budget is only modest.  An important
result of this study is that the often neglected thick disc white
dwarfs contribute a third of the local white dwarf space density.

Neglect of the thick disc was also responsible for a gross
overestimate by \citeasnoun{O.H.D2001} of the Galactic white dwarf mass.  These
authors aimed at classifying halo white dwarfs, but their sample contained a
large fraction of thick disc white dwarfs. Extrapolation of their local numbers
to the vast volume of the Galactic halo resulted in a very large
estimate of the mass. The study presented here suggests a different
interpretation with a more modest contribution of white dwarfs.

Open questions remain. Two scenarios were proposed to explain local
white dwarfs with kinematical properties of the Galactic halo as thin
disc stars, which were accelerated to high velocities after the
companion exploded in a supernova type II \cite{D.K.R2002} or in a
single-degenerate supernova type Ia \cite{Han2003}. Could this be true
for the kinematically identified thick disc and halo white dwarfs in
the SPY sample? The thin disc scenarios make two predictions: 1) the
high-velocity white dwarfs should be single and 2) the masses should
be similar to typical masses of thin disc white dwarfs. Although SPY
is a radial velocity survey for binary white dwarfs, the first check
can not be applied to the \citeasnoun{P.N.H2006} sample, because
suspected binary white dwarfs were intentionally excluded from the
analysis. However, white dwarf masses are known from model atmosphere
analysis of the SPY spectra \cite{K.N.C2001,V.K2005}.  Since star
formation in the thick disc and halo ceased more than 10\,Gyr ago,
young white dwarfs of these populations have low mass progenitors and
are thus expected to have low masses. Thick disc and halo candidates
from the \citeasnoun{P.N.H2006} sample have average masses of
$0.48M_\odot$ and $0.49M_\odot$, respectively, well below the SPY
average of $0.57M_\odot$ \cite{V.K2005}. Thus there is little sign for
a thin disc contamination of this sample. We are in the process of
starting a deep photometric survey, which will trace the white dwarf
populations to large distances from the Galactic plane and will allow
direct measurements of space densities and scale heights.

\ack
The author gratefully acknowledges support by a PPARC/STFC Advanced 
Fellowship.

\section*{References}


\end{document}